\documentclass[journal]{IEEEtran}
\usepackage{graphicx}
\usepackage[hypertex]{hyperref}
\usepackage{subfigure}
\hypersetup{
    colorlinks,%
    citecolor=black,%
    filecolor=black,%
    linkcolor=black,%
    urlcolor=blue
}

\makeatletter
\renewcommand{\@thesubfigure}{\Large(\alph{subfigure})}
\renewcommand{\p@subfigure}{figure\space}
\renewcommand{\p@figure}{figure\space}
\makeatother %

\begin{document}

\title{Progress in characterization of the Photomultiplier Tubes for XENON1T Dark Matter Experiment.}
\author{Alexey~Lyashenko$^a$$^*$, XENON Collaboration
\thanks{$^*$ Corresponding author. E-mail:lyashen@physics.ucla.edu}
\thanks{$^a$ University of California Los Angeles, Department of Physics and Astronomy, 475 Portola Plaza, Los Angeles, CA 90095}%
}
\maketitle
\pagestyle{empty}
\thispagestyle{empty}

\begin{abstract}
We report on the progress in characterization of the Hamamatsu model R11410-21 Photomultiplier tubes (PMTs) for XENON1T dark matter experiment. The absolute quantum efficiency (QE) of the PMT was measured at low temperatures down to -110 $^0$C (a typical the PMT operation temperature in liquid xenon detectors) in a spectral range from 154.5 nm to 400 nm. At -110 $^0$C the absolute QE increased by 10-15\% at 175 nm compared to that measured at room temperature. A new low power consumption, low radioactivity voltage divider for the PMTs is being developed. The measurement results showed that the PMT with the current version of the divider demonstrated a linear response (within 5\%) down to 5$\cdot$10$^4$ photoelectrons at a rate of 200 Hz. The radioactive contamination induced by the PMT and the PMT voltage divider materials satisfies the requirements for XENON1T detector not to exceed a total radioactive contamination in the detector of 0.5 evts/year/1tonn. Most of the PMTs received from the manufacturer showed a high quantum efficiency exceeding 30\%. In the mass production tests the measurements at room temperature showed clear single photoelectron peaks for all PMTs been under study. The optimal operation conditions were found at a gain of 2$\cdot$10$^6$. The operation stability for most of the PMTs was also demonstrated at a temperature of -100 $^0$C. A dedicated setup was built for testing the PMTs in liquid xenon using the XENON1T signal readout components including voltage dividers, cables and feedthroughs. The PMTs tested in liquid xenon demonstrated a stable operation for a time period of more than 5 months.
\end{abstract}


\section{Introduction}
\IEEEPARstart{T}{he} new Hamamatsu model R11410-21 Photomultiplier Tube (PMT) specially designed for XENON1T direct dark matter search experiment \cite{aprile:12} that is being under construction in Gran Sasso National Laboratory (Italy) since June 2013. The detector sensitive volume will accommodate 2.2 ton of liquid xenon monitored by 248 PMTs. The PMT response to various light signals produced in a liquid xenon detector should be well understood as it directly affects the experimental results. Before instrumenting the detector with the PMTs each individual PMT has to be carefully characterized by measuring its single photoelectron peak, gain and afterpulse rate. Some properties like linearity of the response and photocathode uniformity are common for most of the PMTs. These properties can be measured for few PMT samples. Gain, dark count rate, linearity of the response, photocathode uniformity and afterpulsing for the older version R11410-10 PMTs were previously measured \cite{lung:12}.

Some of the PMT characteristics including Quantum Efficiency (QE), gain and linearity of the response could be temperature dependent and, therefore, should be measured at liquid xenon temperature. The performance of the R11410 PMT in a liquid xenon environment was studied elsewhere \cite{baudis:13} demonstrating the stability of gain for a period of more then 5 months and a stable PMT operation in the vicinity of a strong electric field.

Understanding the internal radioactive background in XENON1T detector is a crucial part of the experiment as it will define its sensitivity to the dark matter signal. As shown in \cite{aprile:11} and \cite{aprile:13}, the most of the gamma background and a significant part of the neutron background comes from the PMTs as they contain much more radioactive contaminants than the other elements of the detector. Therefore, it is also crucial to measure the radioactive background from the detector components and the PMTs in particular \cite{aprile:15}.

In the present article we summarize our ongoing campaign on the characterization of the R11410 PMTs that will be employed in XENON1T detector. We report on the recent measurements of the absolute QE at liquid xenon temperature. The measurement results of the linearity of the PMT response are also shown. We present the experimental setups for the mass production tests both at a room temperature and at -100 $^0$C. We also present the experimental setup and the results of the PMT measurements in liquid xenon environment.

\section{PMT arrangement in XENON1T detector}

\begin{figure}[tbp] 
\begin{center}%
\subfiguretopcaptrue
\subfigure[][] 
{
\label{subfig:x1t_cut}
\includegraphics[width=4cm]{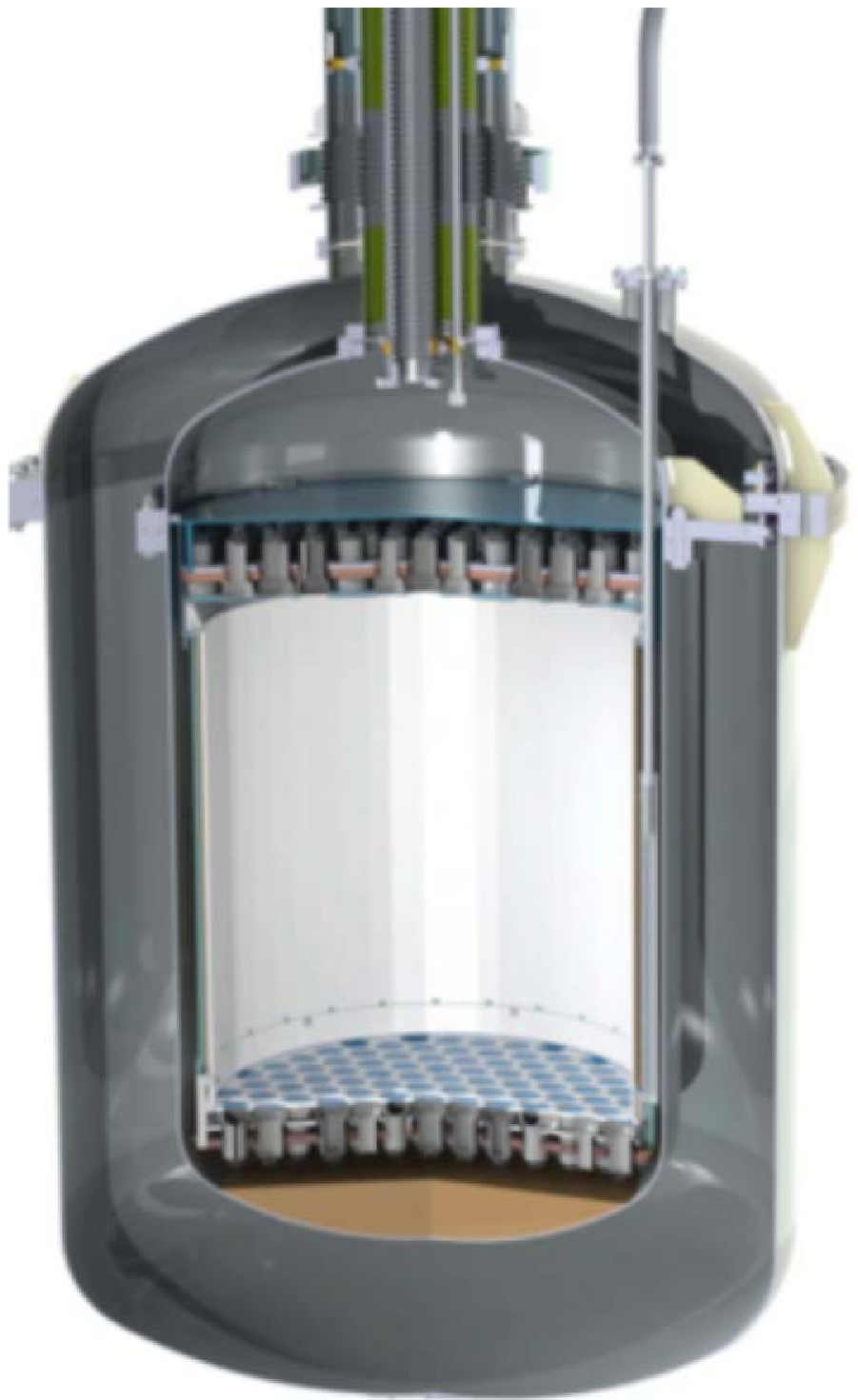}
}
\subfigure[][]
{
\label{subfig:pmt_arrays}
\includegraphics[width=4cm]{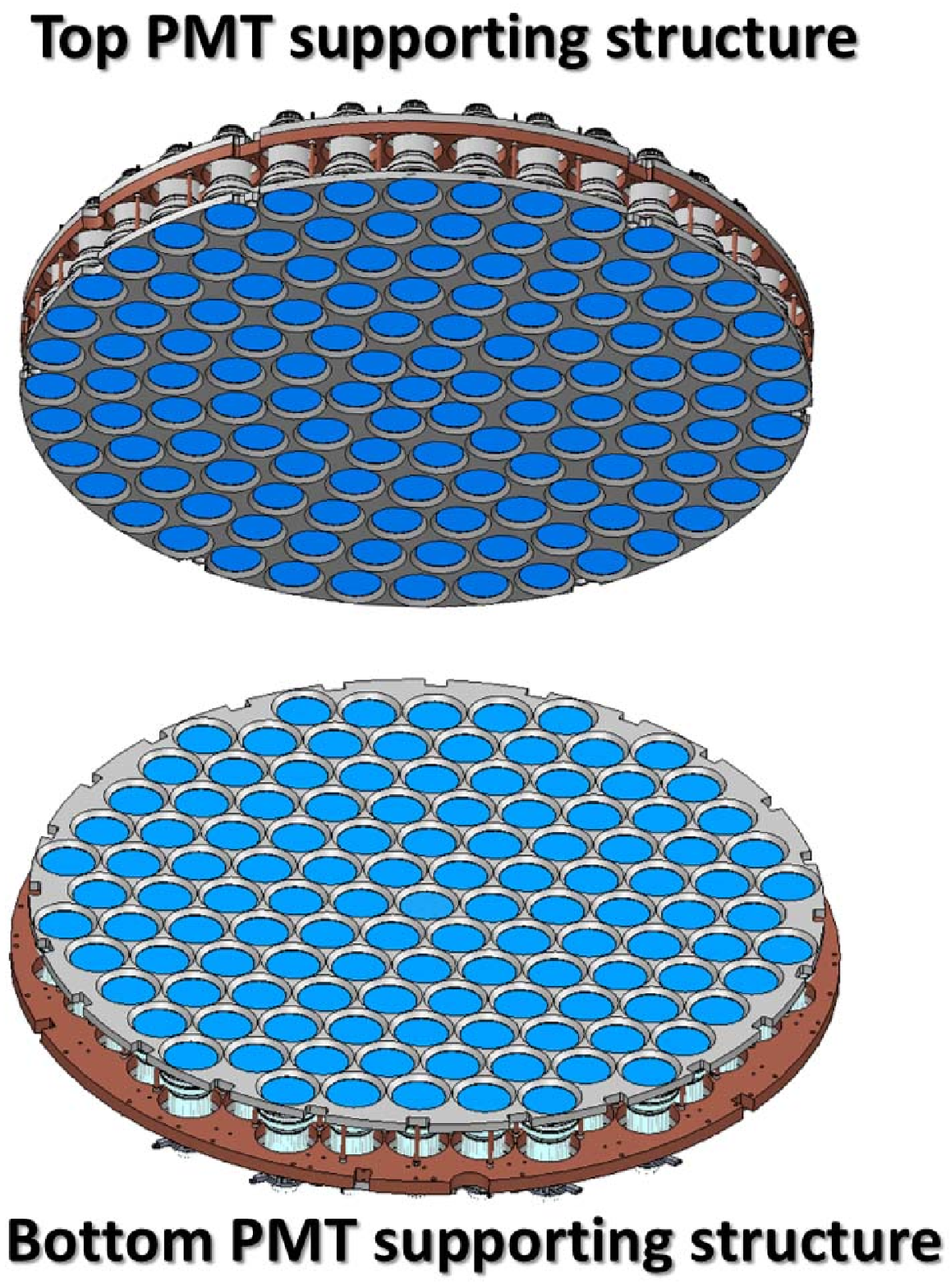}
}
\caption{a) A cross section of XENON1T detector showing its internal structure. b) A 3D rendering of the top and the bottom PMT support structures for XENON1T detector. The bottom PMT array will accommodate 121 PMTs arranged in a hexagonal pattern; the top PMT array will be comprised of 127 PMTs arranged in a circular pattern to improve the resolution of radial event position reconstruction.}
\label{fig:pmt_layout}
\end{center}%
\end{figure}

As mentioned above XENON1T dark matter experiments will accommodate a total of 248 PMTs as shown in \ref{fig:pmt_layout}.

These PMTs will be split between the top and the bottom PMT supporting structures shown in \ref{subfig:pmt_arrays}. The top PMT supporting structure (see \ref{subfig:pmt_arrays}) will be instrumented with 127 PMTs arranged in concentric circles in order to improve the resolution of the event position reconstruction. The bottom PMT array (see \ref{subfig:pmt_arrays}) will incorporate 121 PMTs arranged in a hexagonal pattern to optimize the optical coverage. The gaps between the PMTs in the top and in the bottom PMT arrays will be covered with a single piece PTFE reflector to improve the light collection efficiency.

\section{Measurements of the absolute QE for R11410-10 PMTs}

The measurements of the absolute QE of Hamamatsu model R11410-10 PMTs has been recently reported \cite{lyashenko:14}. Although the measurements were performed for the older PMT version, they can also apply for the newer R11410-21 PMT as it uses the same type of the photocathode.

\begin{figure}[tbp] 
\begin{center}%
\subfiguretopcaptrue
\subfigure[][] 
{
\label{subfig:qe_lt_ka07}
\includegraphics[width=9cm]{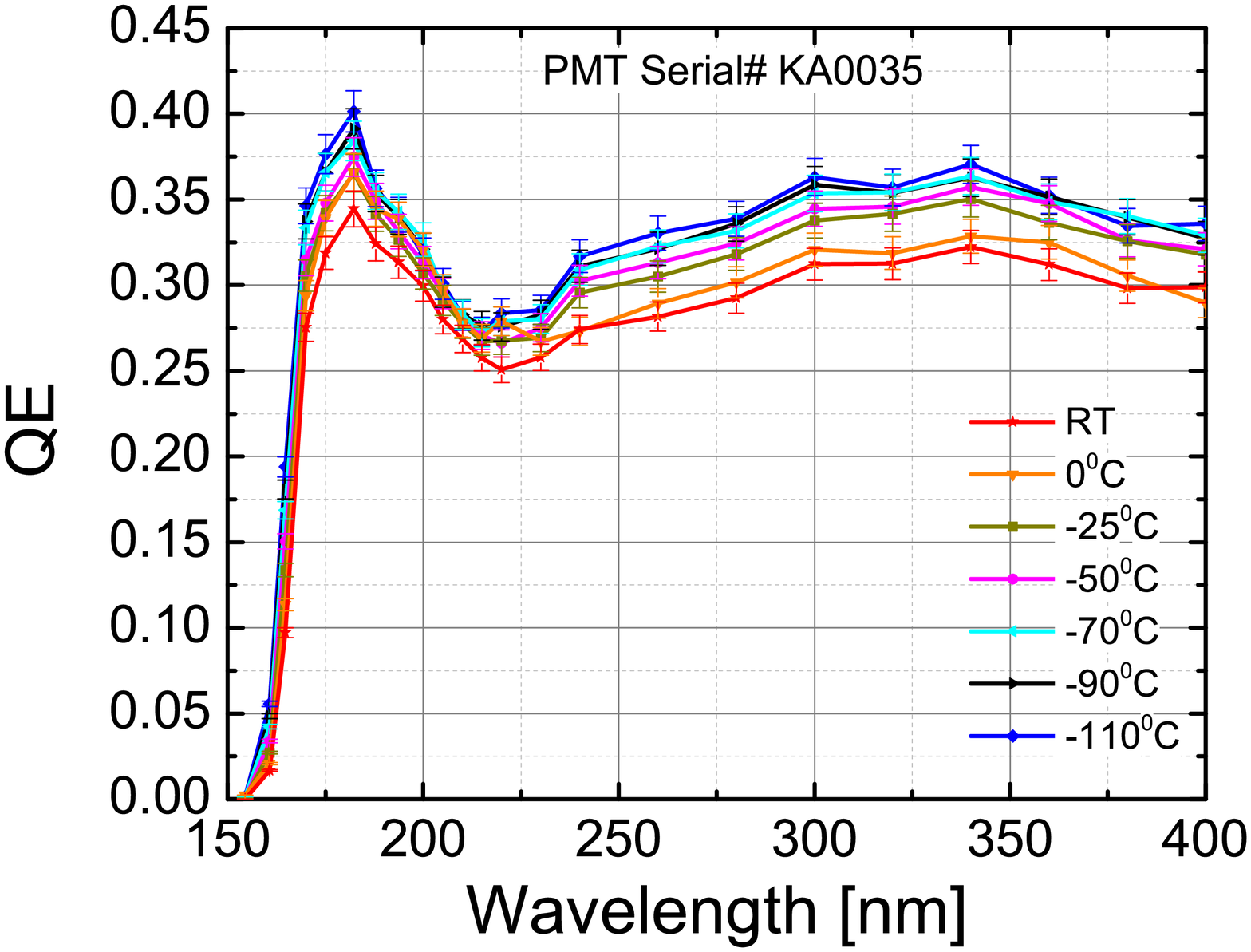}
}
\subfigure[][]
{
\label{subfig:qe_lt_ka35}
\includegraphics[width=9cm]{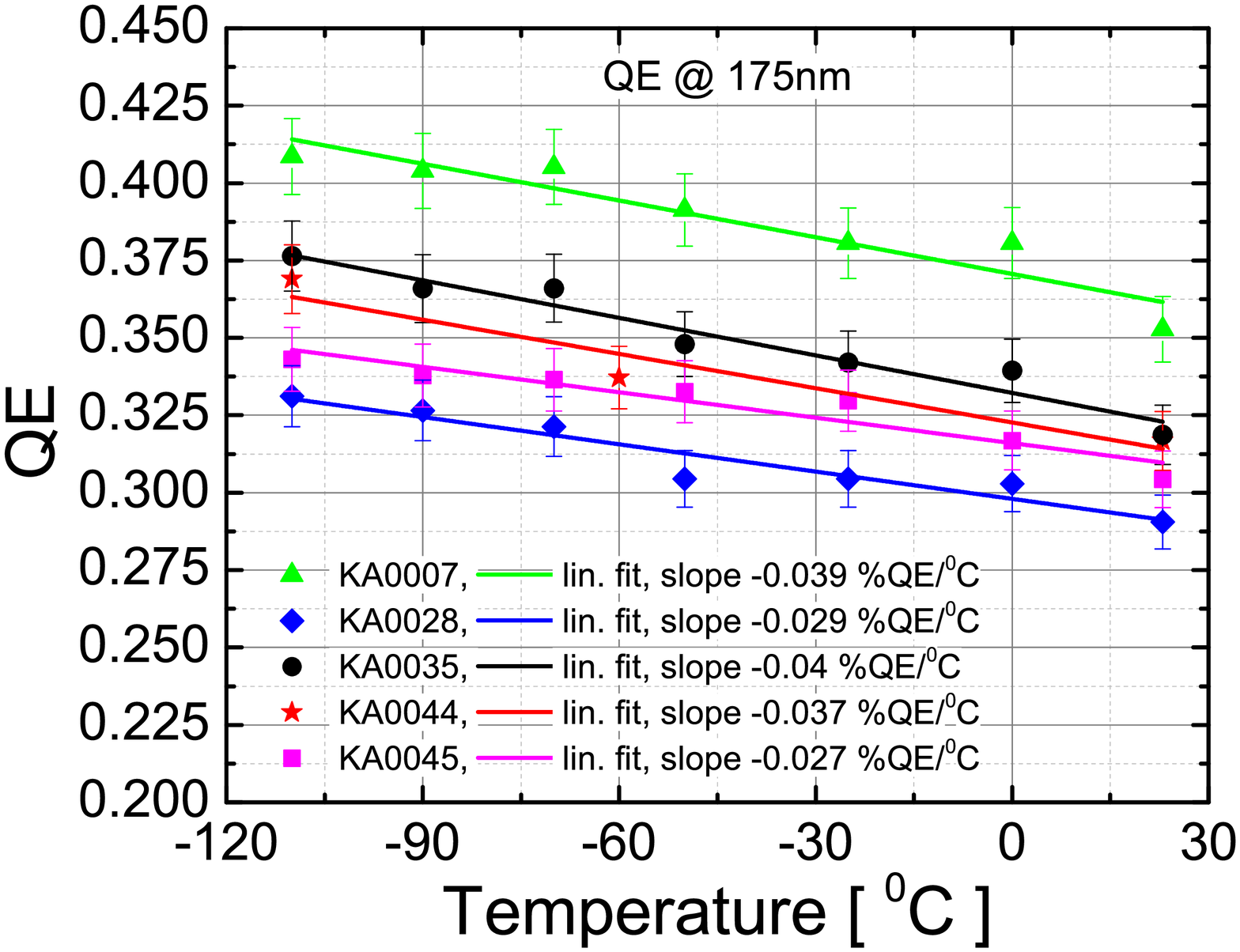}
}
\caption{a) The absolute QE as a function of wavelength measured at various temperatures \protect\cite{lyashenko:14}. The absolute QE was recorded at the following temperatures: Room Temperature (red stars), 0 $^{0}$C (orange triangles), -25 $^{0}$C (dark yellow squares), -50 $^{0}$C (magenta circles), -70 $^{0}$C (cyan left arrows), -90 $^{0}$C (black right arrows) and -110 $^{0}$C (blue diamonds). b) Absolute QE as a function of temperature at a wavelength of 175 nm measured for the five Hamamatsu model R11410-10 PMTs. Solid lines represent linear fits to each of the PMTs.}
\label{fig:qe_lowt}
\end{center}%
\end{figure}

The absolute QE of five Hamamatsu model R11410-10 PMTs was measured at low temperatures down to -110 $^{0}$C in a spectral range from 154.5 nm to 400 nm. As shown in \ref{fig:qe_lowt} that during the PMT cooldown from room temperature to -110 $^{0}$C (operation temperature of the PMTs in liquid xenon detectors) the QE increases by 10-15\% at 175 nm. The increase of the QE at low temperatures can be explained as follows. In the photocathode bulk a dominant energy loss mechanism for a photoelectron prior its emission into vacuum is the collisions with optical phonons. A decrease of the phonon-photoelectron cross-section \cite{araujo:03} with temperature results in a reduced energy loss by the photoelectron and, therefore, leads to an improved quantum efficiency. The increase of QE of the PMT at low temperature will result in the improved single photon sensitivity. This will allow for lowering of the energy threshold and therefore, improving the detector sensitivity to the low energy WIMPs.

\section{PMT voltage divider}

The purpose of the PMT voltage divider is to supply proper bias voltages (recommended by the manufacturer) for each of the PMT dynodes and to pick up the signal from the PMT anode. Usually it is realized using a passive resistive linear circuit. In the design of the PMT voltage divider for the low background liquid noble gas detectors such as XENON1T, several important requirements have to be met. First, it is a low power dissipation of the divider. A constant current flowing through the resistive elements of the divider will produce heat that could cause a formation of bubbles or initiate boiling when the divider operates in a noble liquid. These processes have to be avoided as it can affect the stability of the detector by causing some sparking issues in the regions of a high electric field in the detector. In order to avoid the formation of the bubbles in the liquid xenon the heat flux (defined as an electric power per surface area) produced in the divider should not exceed $2\cdot10^4$ W/m$^2$ as shown in \cite{haruyama:02}. Second, the voltage divider has to be made of low radioactivity materials. As indicated in \cite{aprile:11} and \cite{aprile:13} the components of the PMT divider could make a considerable contribution to the radioactive background. And third, it is a stable operation at low temperatures.

The PMT voltage divider for XENON1T detector is being designed to fulfill the above requirements. The electrical scheme of the divider will be presented in the next technical paper by XENON1T collaboration.

\begin{figure}[tbp] 
\begin{center}%
\includegraphics[width=4cm]{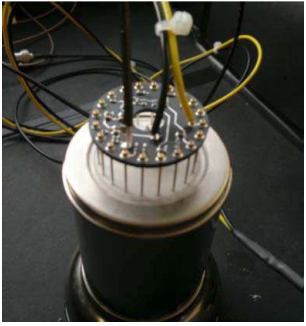}
\caption{A photograph of the voltage divider PCB mounted on the PMT.}
\label{fig:pmt_base_photo}
\end{center}%
\end{figure}

The Printed Circuit Board (PCB) of the divider was made of Cirlex\footnote{\href{http://www.cirlex.com/}{Cirlex} is a registered trademark of DUPONT} polyimide laminate known to have significantly lower radioactivity compared to the traditional PCB materials. A photograph of the Cirlex PCB of the divider mounted on the PMT is shown in \ref{fig:pmt_base_photo}. The divider is characterised by a very low power consumption of 0.024 W per PMT at -HV = 1500 V (a nominal PMT operation voltage). The heat flux through the surface area of the resistors in the PMT divider can be estimated as 0.024 W / (14 $\cdot$ 2 mm $\cdot$ 1.25 mm) = 760 W/m$^2$, where 14 is the number of Mega Ohm - scale resistors in the divider and 2 mm $\cdot$ 1.25 mm is the surface area of the each Mega Ohm - scale resistor. According to \cite{haruyama:02} this heat flux is low enough not to form any bubbles in the liquid xenon.

\begin{figure}[tbp] 
\begin{center}%
\includegraphics[width=9cm]{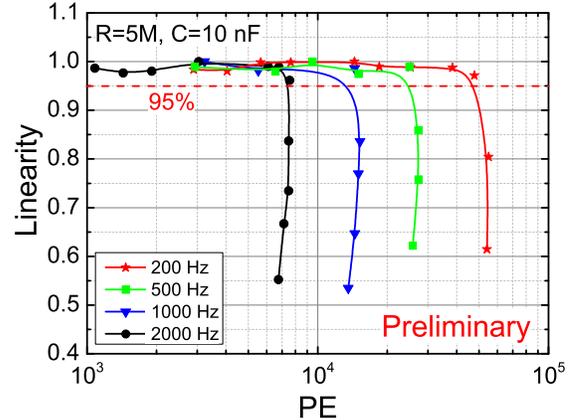}
\caption{Preliminary results of the PMT linearity measurements. The linearity of the PMT response to 1 $\mu$S wide light pulses as a function of the number of photoelectrons per pulse measured at various pulse repetition rates. The PMT gain is $4.85\cdot10^6$.}
\label{fig:linearity}
\end{center}%
\end{figure}

The linearity of the PMT response to the light pulses could also be affected by the voltage divider as well as by the avalanche charge distribution inside the PMT. When an intense light pulse enters a PMT a large current flows in the latter dynode stages increasing the space charge density and causing current saturation. The space charge effects depend on the electric field distribution and intensity between each dynode. The large current in the last dynode stages could cause a voltage drop at the resistors of the PMT voltage divider thus slightly redistributing the voltages applied to the dynodes and affecting the linearity of the PMT response to the light pulses. The linearity of the PMT divider could be improved by adding capacitors at the last amplification stages. The linearity of the PMT response to 1 $\mu$S wide (an approximate pulse width of S2 signals in XENON1T detector) light pulses as a function of the number of photoelectrons per pulse was measured as described in \cite{hama}. It was measured at a pulse repetition rate of 200 Hz, 500 Hz, 1000 Hz and 2000 Hz as shown in \ref{fig:linearity}. The measurements were performed at a PMT gain of $4.85\cdot10^6$.

As seen in \ref{fig:linearity}, the PMT with the voltage divider shows a linear response within a deviation of 5\% up to $5\cdot10^4$, $2.5\cdot10^4$, $1.5\cdot10^4$ and $7.5\cdot10^3$  photoelectrons in the light pulse at a rate of 200 Hz, 500 Hz, 1000 Hz and 2000 Hz respectively.

A detailed description of the main design concepts in the divider development for R11410 PMTs as well as a comprehensive survey of the linearity measurements can be found elsewhere \cite{behrens:13}.

\section{Radioactive screening}

We employed several methods to assess the radioactive contamination of the PMT materials including the High Purity Germanium (HPGe) underground detectors and Glow Discharge Mass Spectrometry (GDMS) \cite{aprile:11:1}. Each of the material used in the PMT was studied using the above mentioned techniques. The measured radioactive contaminations by various radioactive isotopes present in the PMT materials fulfill the requirements for XENON1T detector with a total radioactive background budget of 0.5 evts/year/1T. A comprehensive radioactive screening report for the PMTs and the PMT materials will be soon published by XENON collaboration.

\section{Mass production tests}

\begin{figure}[tbp] 
\begin{center}%
\includegraphics[width=7cm]{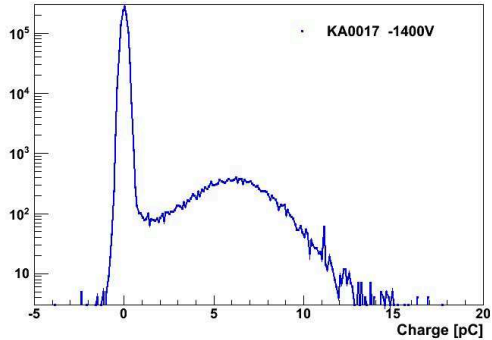}
\caption{An example of the single photoelectron spectrum recorder at a bias voltage of -1400V corresponding to a gain of about $2\cdot10^6$}
\label{fig:spe}
\end{center}%
\end{figure}

So far we have received 220 PMTs from the manufacturer. Most of the PMTs showed rather high QE exceeding 30\%. For the mass production tests at room temperature we adopted the experimental PMT test facility at Max-Planck-Institut f\"{u}r Kernphysik (MPIK) at Heidelberg. It allows for simultaneous calibration of 12 PMTs by measuring the single photoelectron response, transit time spread, after pulse rate and spectrum, and high voltage calibration. An example of a single photoelectron spectrum recorded using this setup is shown in \ref{fig:spe}. The PMT was biased at -1400V corresponding to a gain of about $2\cdot10^6$.

In a real experiment one should operate PMTs at a lowest possible bias voltage to avoid possible sparking issues. At the same time this bias voltage has to provide high enough gain to ensure clear SPE peak. A single photoelectron (SPE) spectrum was recorded for all PMTs. These spectra were then analysed to extract the peak-to-valley (P/V) ratios, the PMT amplification factors (Gains) and the SPE resolution. We learned from these measurements that the optimal PMT gain to provide low enough PMT operation voltage, a satisfactory peak-to-valley ratio of about 3.5 and the SPE resolution of about 30\% is around $2\cdot10^6$.

A dedicated setup at Max-Planck-Institut f\"{u}r Kernphysik (MPIK) at Heidelberg for the mass production tests in cold. In this setup 12 PMTs can be cooled down simultaneously with a cold nitrogen gas in order to study the PMT characteristics in cold and to subject the PMTs to a thermal stress test. The gas is cooled by convection through the constant flow of liquid nitrogen in the copper coil. The measurements the dark count rate (noise PMT signals induced by a number of sources including field emission, leakage current, and thermionic emission) at -100 $^0$C were performed for all PMTs. A vast majority of the PMTs demonstrated a stable low temperature operation. According to the manufacturer's specifications the rate of dark counts for these type of the PMTs should not exceed 200 Hz. It has to be mentioned that each of the tested PMTs were cooled down for at least 3 times.

\section{Measurements in liquid xenon}

\begin{figure}[tbp] 
\begin{center}%
\includegraphics[width=9cm]{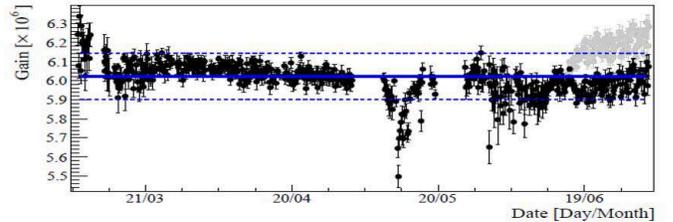}
\caption{Long-term stability of the gain of a R11410 immersed in LXe, as measured over a period of about
5 months \protect\cite{baudis:13}. The gain is stable within 2\%, as indicated by the dashed lines. Periods in which the gain shows larger variations can be correlated to changing experimental conditions. At the end of the run, the gray points show the measurements while the corresponding black points are corrected and take into account a sudden change in the LXe pressure and temperature. A similar correction cannot be done reliably for the short spikes in the gain measurements, hence they are presented as measured.}
\label{fig:gain_vs_time}
\end{center}%
\end{figure}

A dedicated experimental setup for performing the PMT tests in liquid xenon was built at University of Zurich. The setup comprises of a cryostat that accommodates 5 PMTs to be submerged in liquid xenon. The PMTs biasing and the signal readout are realized using the voltage dividers, cables and cable feedthroughs that are identical to those that will be used in XENON1T detector. In \ref{fig:gain_vs_time} we present the gain stability measurements in liquid xenon. The gain of a R11410-21 PMT immersed in LXe was monitored for a period of 5 months.

As we learned from \ref{fig:gain_vs_time}, the gain is stable within 2\%, as indicated by the dashed lines. Periods in which the gain shows larger variations can be correlated to changing experimental conditions. The gray points on the top right of the \ref{fig:gain_vs_time} show the measurements while the corresponding black points are corrected and take into account a sudden change in the LXe pressure and temperature. A similar correction cannot be done reliably for the short spikes in the gain measurements, hence they are presented as measured.

\section{Conclusions}

XENON1T will incorporate 248 Hamamatsu model R11410-21 Photomultiplier tubes (PMTs) shared between the top and bottom PMT arrays. A comprehensive campaign on the characterization of these PMTs and on the study of some important PMT properties is in full swing.

The absolute quantum efficiency (QE) of the PMT was measured at low temperatures down to -110 $^0$C (a typical the PMT operation temperature in liquid xenon detectors) in a spectral range from 154.5 nm to 400 nm. At -110 $^0$C the absolute QE increased by 10-15\% at 175 nm compared to that measured at room temperature.

We developed a new low power consumption, low radioactivity voltage divider for the PMTs. It is characterised by a very low power consumption of 0.024 W per PMT at -HV = 1500 V (a nominal PMT operation voltage) to avoid the formation of bubbles in the liquid xenon. The preliminary results of the linearity measurement for the current version of the divider showed that a PMT with this voltage divider shows a linear response up to $5\cdot10^4$, $2.5\cdot10^4$, $1.5\cdot10^4$ and $7.5\cdot10^3$  photoelectrons in the light pulse at a rate of 200 Hz, 500 Hz, 1000 Hz and 2000 Hz correspondingly.

The measured radioactive contamination induced by the PMTs and the PMT voltage divider materials fulfill the requirements of XENON1T detector to not exceed the total background of 0.5 evts/year/1tonn.

220 PMTs received from Hamamamtsu Photonics K.K. showed an average QE exceeding 30\% at 175 nm at room temperature. At the room temperature, we measured on a dedicated setup the single photoelectron peaks for all deliverd PMTs showing that the optimal PMT gain to provide the lowest operation voltage, ensuring satisfactory peak-to-valley ratio and SPE resolution is around $2\cdot10^6$. The operation stability for most of the PMTs was also demonstrated at a temperature of -100 $^0$C.

A dedicated setup was built for testing the PMTs in liquid xenon. The PMTs tested in liquid xenon demonstrated a stable operation for over a 5 months period.

\section*{Acknowledgment}
We gratefully acknowledge support from NSF, DOE, SNF, Volkswagen Foundation, FCT, Region des Pays de la Loire, STCSM, NSFC, DFG, MPG, Stichting voor Fundamenteel Onderzoek der Materie (FOM), the Weizmann Institute of Science, the EMG research center and INFN. We are grateful to LNGS for hosting and supporting XENON1T.

\end{document}